\documentclass{PoS}
\let\ifpdf\relax

\usepackage[authoryear,square]{natbib}
\bibpunct{(}{)}{;}{a}{}{,}

\title{Magnetic Field Tomography in Nearby Galaxies with the Square Kilometre Array}

\ShortTitle{Magnetic Field Tomography in Nearby Galaxies with the SKA}

\author{\speaker{George Heald}$^{1,2}$\thanks{on behalf of the SKA Cosmic Magnetism Working Group}, Rainer Beck$^3$, W.~J.~G. de Blok$^{1,4,2}$, Ralf-J\"urgen Dettmar$^5$, Andrew Fletcher$^6$, Bryan Gaensler$^7$, Marijke Haverkorn$^{8,9}$, Volker Heesen$^{10}$, Cathy Horellou$^{11}$, Marita Krause$^3$, Sui Ann Mao$^3$, Niels Oppermann$^{12}$, Anna Scaife$^{10}$, Dmitry Sokoloff$^{13}$, Jeroen Stil$^{14}$, Fatemeh Tabatabaei$^{15}$, Keitaro Takahashi$^{16}$, Russ Taylor$^{14,4}$ and Anna Williams$^{17}$\\
$^1$ASTRON, Postbus 2, 7990 AA Dwingeloo, the Netherlands;
$^2$Kapteyn Astronomical Institute, PO Box 800, 9700 AV Groningen, the Netherlands;
$^3$Max-Planck-Institut f\"ur Radioastronomie, Auf dem H\"ugel 69, 53121 Bonn, Germany;
$^4$University of Cape Town, Private Bag X3, Rondebosch 7701, South Africa;
$^5$Ruhr-Universit\"at Bochum, Universitaetsstrasse 150, 44780, Bochum, Germany;
$^6$School of Mathematics and Statistics, Newcastle University, Newcastle upon Tyne NE1 7RU, UK;
$^7$The University of Sydney, Sydney, NSW 2006, Australia;
$^8$Department of Astrophysics/IMAPP, Radboud University Nijmegen, P.O. Box 9010, 6500 GL Nijmegen, The
Netherlands;
$^9$Leiden Observatory, Leiden University, P.O. Box 9513, 2300 RA Leiden, The Netherlands;
$^{10}$School of Physics and Astronomy, University of Southampton, Southampton SO17 1BJ, UK;
$^{11}$Chalmers University of Technology, Onsala Space Observatory, 43992, Onsala, Sweden;
$^{12}$Canadian Institute for Theoretical Astrophysics, University of Toronto, 60 St. George Street, Toronto ON, M5S 3H8, Canada;
$^{13}$Department of Physics, Moscow University, 119992 Moscow, Russia;
$^{14}$The University of Calgary, Calgary, AB T2N 1N4, Canada;
$^{15}$Max-Planck-Institut f\"ur Astronomie, K\"onigstuhl 17, 69117 Heidelberg, Germany;
$^{16}$Kumamoto University, 2-39-1 Kurokami, Kumamoto 860-8555, Japan;
$^{17}$Department of Astronomy, University of Wisconsin-Madison, Madison, WI, USA\\
E-mail: \email{heald@astron.nl}
}

\abstract{Magnetic fields play an important role in shaping the structure and evolution of the interstellar medium (ISM) of galaxies, but the details of this relationship remain unclear. With SKA1, the 3D structure of galactic magnetic fields and its connection to star formation will be revealed. A highly sensitive probe of the internal structure of the magnetoionized ISM is the partial depolarization of synchrotron radiation from inside the volume. Different configurations of magnetic field and ionized gas within the resolution element of the telescope lead to frequency-dependent changes in the observed degree of polarization. The results of spectro-polarimetric observations are tied to physical structure in the ISM through comparison with detailed modeling, supplemented with the use of new analysis techniques that are being actively developed and studied within the community such as Rotation Measure Synthesis. The SKA will enable this field to come into its own and begin the study of the detailed structure of the magnetized ISM in a sample of nearby galaxies, thanks to its extraordinary wideband capabilities coupled with the combination of excellent surface brightness sensitivity and angular resolution.}

\FullConference{
Advancing Astrophysics with the Square Kilometre Array\\
June 8-13, 2014\\
Giardini Naxos, Sicily, Italy}

\newcommand{\HI}{{\sc H\,i}}
\newcommand{\rmclean}{{\tt RMCLEAN}}
\newcommand{\radm}{$\mathrm{rad\,m^{-2}}$}

\begin{document}

\section{Overview and Techniques}

%

A key issue to be addressed with the SKA is the three-dimensional structure of magnetic fields in galaxies. One of the most powerful tools for studying these structures is the Faraday rotation of polarized synchrotron radiation. Faraday rotation, caused by the birefringence of magnetoionized plasma, rotates the plane of linear polarization by an angle proportional to $\lambda^2$, where the constant of proportionality is called the rotation measure (RM). In the simplest cases the RM is equal to the ``Faraday depth'', the integral of the thermal electron density weighted line-of-sight magnetic field. Modern radio spectropolarimetry typically makes use of the so-called RM Synthesis \citep{brentjens_debruyn_2005} technique to combine broadband, multi-channel observations into Faraday depth spectra, which encode the polarized emission visible as a function of the Faraday depth. The benefits of the RM Synthesis technique are discussed elsewhere \citep[e.g.][]{heald_2009} but include sensitivity to faint Faraday-rotated polarized emission, along with the related ability to recover -- at least partially -- polarized emission originating at a range of Faraday depths within the line of sight (LOS) cylinder. Such LOS are typically called ``Faraday thick'' because their Faraday depth spectra show broad profiles. They are probed by frequencies between approximately 500 MHz and 2 GHz.

RM Synthesis is based on a Fourier transform of the observed complex linear polarization $P(\lambda^2)=Q(\lambda^2)+iU(\lambda^2)$ from the $\lambda^2$ domain into the Faraday depth ($\phi$) dimension, where $Q$ and $U$ are the linear Stokes parameters. Since the $\lambda^2$ coverage of any given observation is incomplete, there is a sampling function called the RM Spread Function (RMSF) in the Faraday domain. The RMSF can be deconvolved from the Faraday spectrum through the use of an algorithm called \rmclean\ \citep{heald_etal_2009}, but it has been recognized \citep[e.g.][]{farnsworth_etal_2011} that the technique has some drawbacks. It is usually complemented by fitting in the $(Q,U)$ domain \citep[e.g.][]{osullivan_etal_2012}, which requires \emph{a priori} knowledge or assumptions about the physical nature of the emitting region(s), but can lead to a more stable signal reconstruction than deconvolution. In this chapter we assume that data analysis techniques will make use of a suitable combination of RM Synthesis, deconvolution, and guided fitting to recover physical parameters from polarization observations.

SKA observations in Band 4 (Beck et al., this volume) will provide exquisite sensitivity to the small scale structure of magnetic fields in the disks of galaxies and the relation to other morphological information such as the gas distribution. Such observations are rather sensitive because the radiation has suffered very little depolarization, which is caused by structure in the gas and magnetic field distributions on scales smaller than the resolution element of the telescope. Internal depolarization is ubiquitous in galaxies, causes a strongly frequency-dependent modulation of the polarization degree, and carries a great deal of information about the physical conditions of the ISM. With the SKA we can make use of lower frequency observations to provide a new dimension of information that is highly complementary to the higher frequency approach, gaining new insight into the small scale structure of the ISM through detailed modeling of the depolarization. This can then be compared with the small-scale structure of the gaseous component.

In this contribution we seek to outline the applicability of this observing setup and analysis approach to gaining leverage on the structure of the ISM in galaxies. While the focus is on spiral galaxies, we note that other objects can be studied in this way, e.g. AGN and intracluster media.

With SKA1, Faraday tomography can be performed in nearby galaxies with excellent spatial resolution (e.g. $\sim34$~pc for M~83) and with suitable surface brightness sensitivity. Even in the 50\%-sensitivity ``early science'' phase, SKA1 will provide excellent prospects for this project. The SKA will open the door to the study of these phenomena in even finer detail in the nearest targets and will allow study of the evolution of magnetic field structures in galaxies.

\section{Scientific goals}
\label{section:science}

\subsection{Small-scale 3D gas and magnetic field structures}

%

Polarization and depolarization are two sides of the same coin. Long considered as a nuisance to observers, especially at long wavelengths where it usually becomes more severe and affects prospects of detection, depolarization has been traditionally more appreciated by theorists who recognized its potential as a powerful probe of the magneto-ionic medium \citep{burn_1966,sokoloff_etal_1998}. The SKA, with its high sensitivity and broad bandwidth compared to all existing instruments, opens up the possibility to exploit this potential, and learn about both the small-scale and the large-scale structure of the magnetic field in galaxies, and about their interactions with the populations of electrons, thermal and non-thermal (cosmic ray electrons). 

The various depolarization mechanisms acting in galaxies are well understood. Most of them have to do with Faraday rotation (differential Faraday rotation, internal or external Faraday dispersion) and are thus wavelength-dependent and related to the thermal component of the gas and the magnetic field component along the line of sight. Depolarization can also be produced within the observing beam, for example by a tangled magnetic field (in which case the observed degree of polarization is wavelength-independent) or due to the averaging of several sources (in which case it could be wavelength-dependent). The signature of random but anisotropic magnetic fields can also be predicted.

\begin{figure}
\centering
\includegraphics[width=0.51\textwidth]{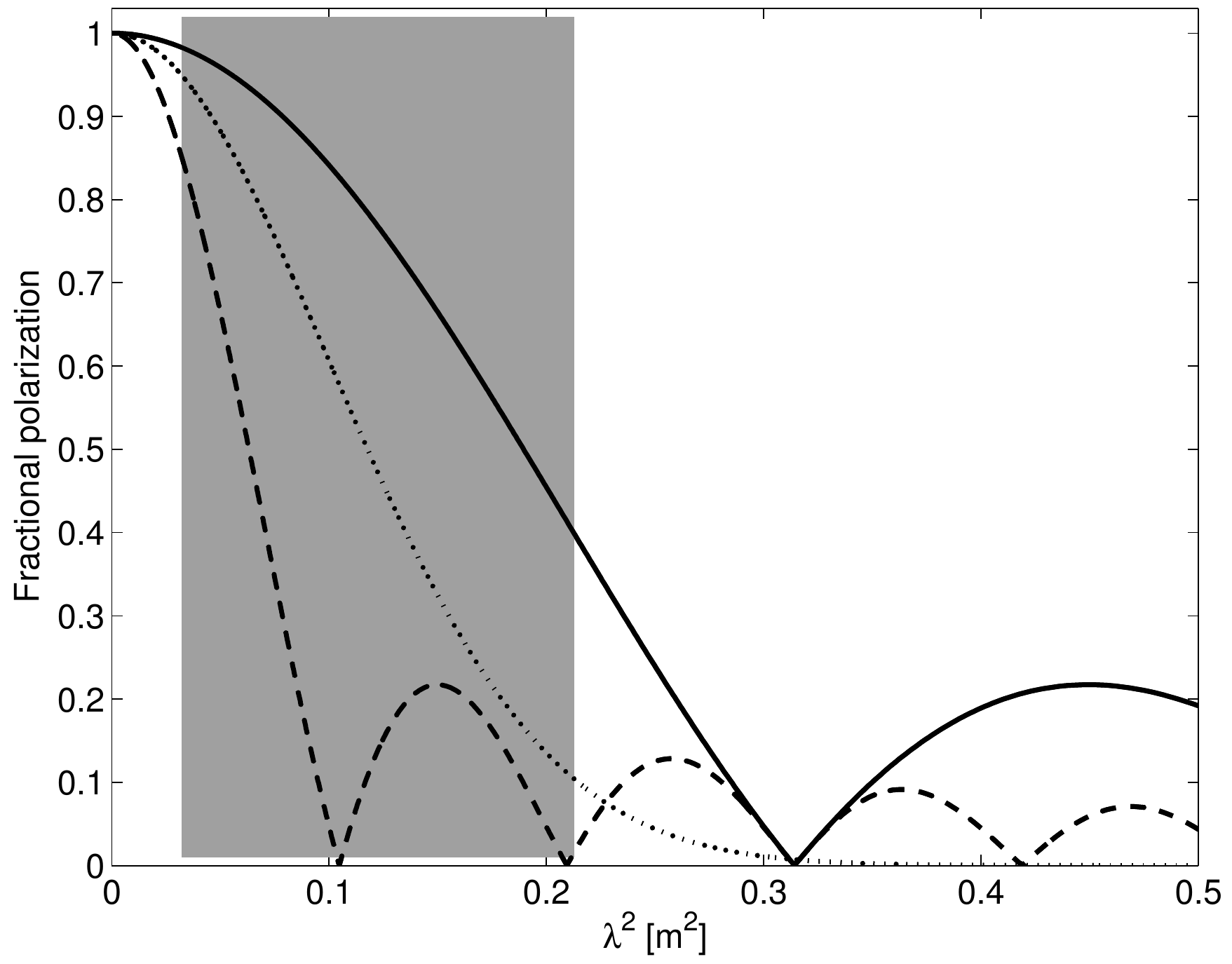}\includegraphics[width=0.5\textwidth]{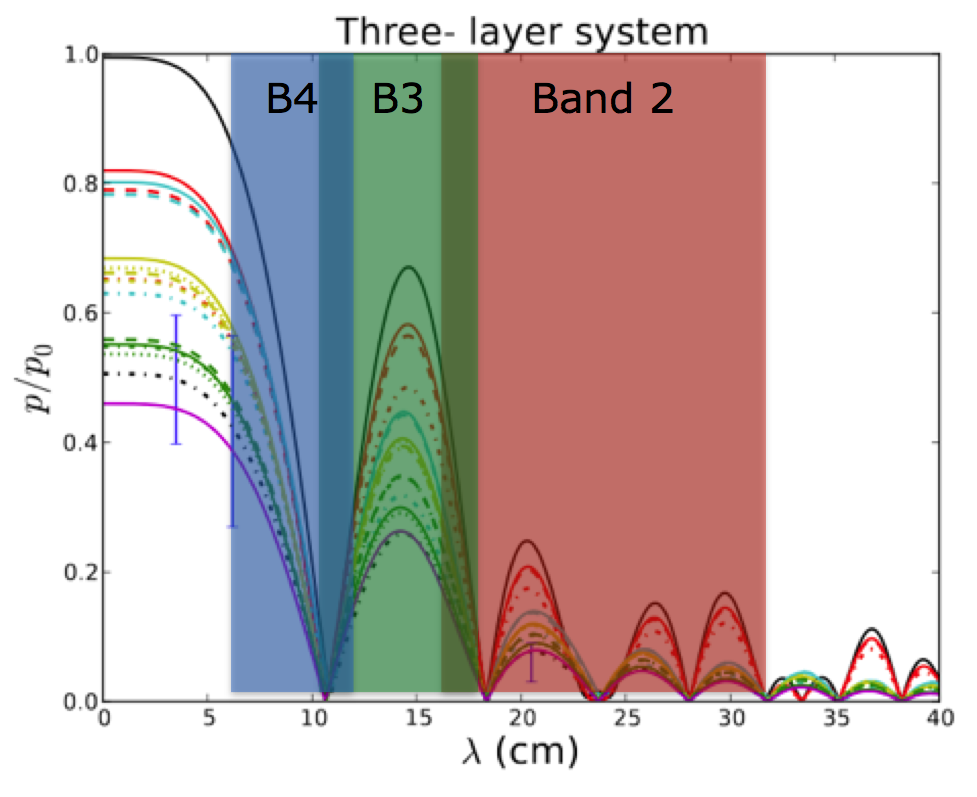}\\
\includegraphics[width=0.55\textwidth]{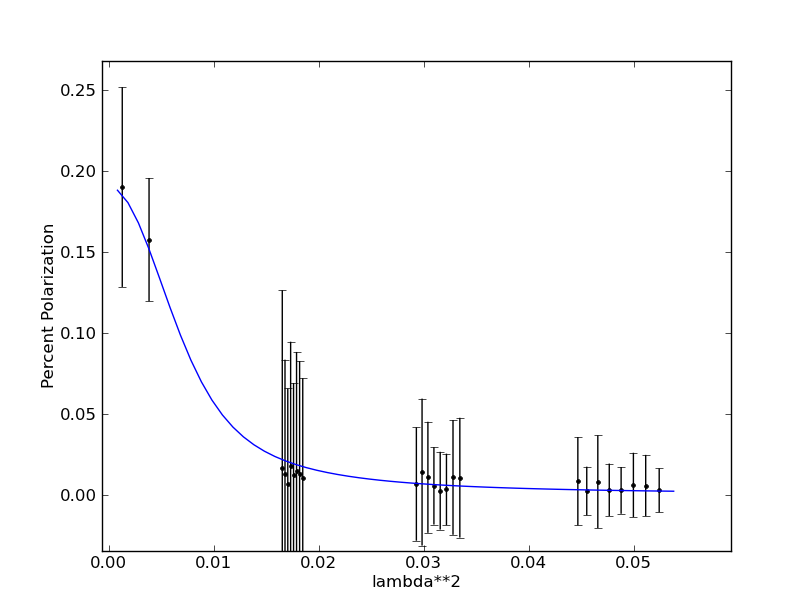}\includegraphics[width=0.45\textwidth]{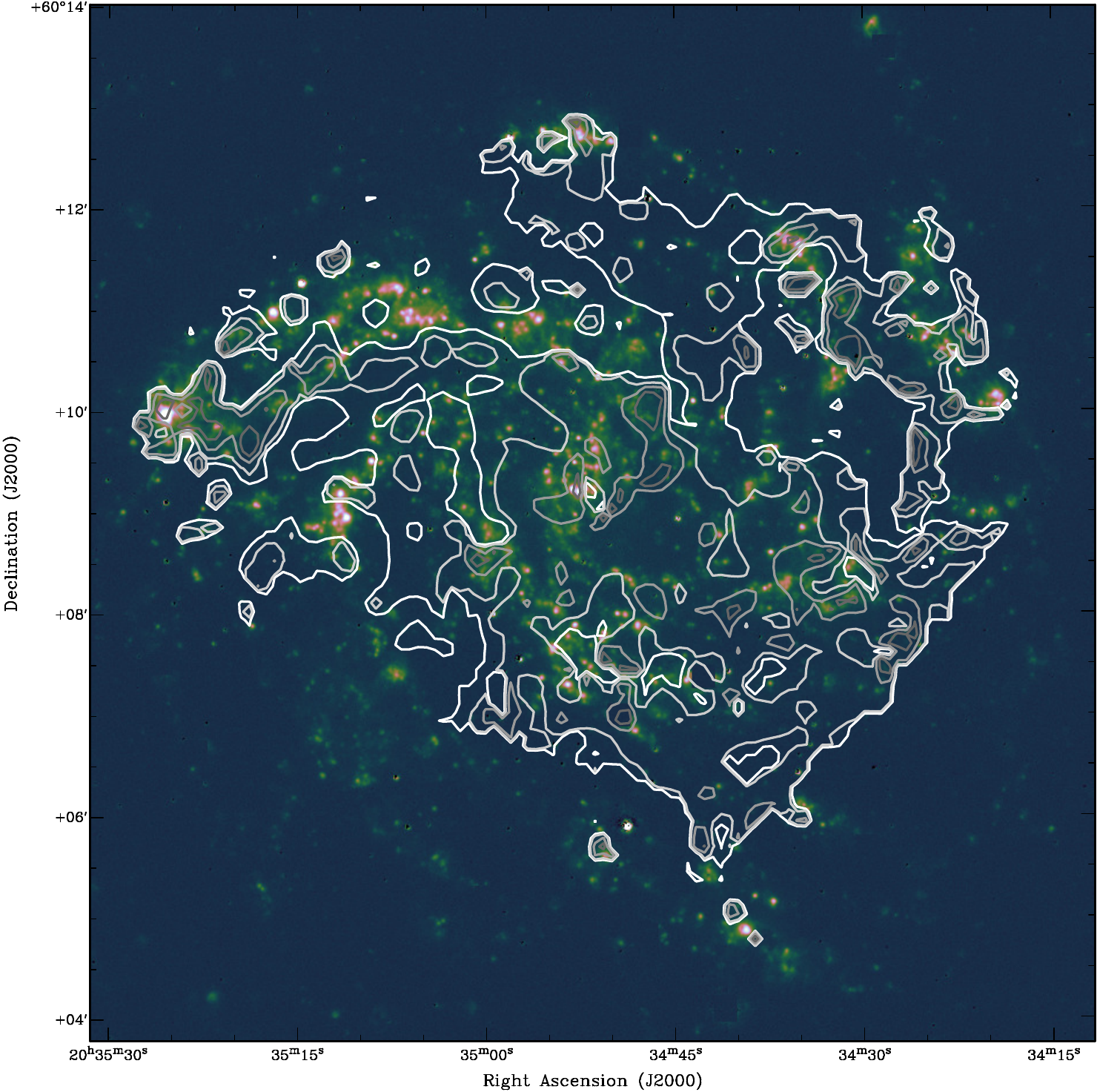}
\caption{{\it Top left}: simple models of Faraday dispersion, using a Burn slab (thick line and dashed line) or a Gaussian distribution (dotted line). For the Burn slab the thicker distribution of Faraday depth corresponds to the dashed line and therefore to stronger depolarization across the bandwidth of interest (gray shaded region).
{\it Top right}: Various depolarization models reproduced from \citet{shneider_etal_2014} and overlaid with colored regions indicating the approximate wavelength coverage of the various SKA1-MID bands.
{\it Bottom left}: Faraday dispersion measured in a particular region of the NE interarm region in NGC~6946 \citep{williams_inprep}. Measurements of the polarized fraction were made at 3,6,13,18, and 22~cm and fitted with a Gaussian model.
{\it Bottom right}: Strength of Faraday dispersion within NGC~6946, determined by fits like the one in the bottom left panel and presented as contours overlaid on the H$\alpha$ image from \citet{ferguson_etal_1998} (H$\alpha$ image on a log stretch). Increased turbulent depolarization tends to be associated with \textsc{H\,ii} regions.
}
\label{figure:depol}
\end{figure}

For a number of reasons, observations in SKA band 2 ($\nu=650-1670$ MHz or $\lambda=18-46$ cm for SKA1-SUR, or $\nu=950-1760$ MHz for SKA1-MID; see Table \ref{table:summary}) are ideal. First, the typically steep synchrotron spectral index ($\alpha\gtrsim0.7$, for $S_\nu\propto\nu^{-\alpha}$) from the ISM in galaxies means that the continuum emission is strong at these moderate radio frequencies. Moreover the frequency range covered in Band 2 is well-suited to measure the variations of the fractional polarization, for the typical ISM properties in spirals. 
In a mixed synchrotron-emitting and Faraday-rotating medium, differential Faraday rotation results in a 
decrease of the fractional polarization with $\lambda^2$, as illustrated in Figure \ref{figure:depol}. 
The broader the Faraday dispersion function (which describes the intrinsic polarized emission as a function of Faraday depth), the faster the decrease. For LOS with narrow distributions of Faraday depth, there is little depolarization across the band as shown for example by \citet{arshakian_beck_2011}.
The variations are shown for  top-hat (``Burn slab'') models of different widths (10~rad~m$^{-2}$, solid line, and 30~rad~m$^{-2}$, dashed line). A Gaussian model with a dispersion of 30~rad~m$^{-2}$ is also shown (dotted line). 
Faraday dispersion by a random field will accentuate this decrease \citep{sokoloff_etal_1998}. In some magnetic field configurations, such as for helical fields along the line of sight, the degree of polarization may actually increase with wavelength \citep{sokoloff_etal_1998,brandenburg_stepanov_2014,horellou_fletcher_2014}. 


As early examples of Faraday tomography in the pre-SKA era, recent sparsely sampled multifrequency polarimetric observations of a small number of nearby galaxies are now providing new insight into the ordered and random magnetic fields in the ISM. For example, a combined observation of NGC~6946 using the Westerbork Synthesis Radio Telescope (WSRT) at $\lambda=13\,\mathrm{cm}$, 18~cm, and 22~cm, together with 3.5~cm and 6.2~cm data \citep[from][]{beck_etal_2007}, is being used to model the depolarization \citep[][see Figure \ref{figure:depol}]{williams_inprep}, probe new Faraday depths with RM Synthesis, and recover detailed estimates of the magnetic field structure across the entire galaxy.  Assuming typical ISM properties and a simple configuration of a large-scale magnetic field along the line of sight and turbulent magnetic cells of equal size and uniform distribution, the Burn slab model \citep{burn_1966,sokoloff_etal_1998} can also be utilized to begin to tease out the coherence length of the large-scale line of sight magnetic field and correlation lengths of line of sight random magnetic fields. Applying this rather simple model to all of the various ISM environments in NGC 6946 is a stepping stone toward robustly estimating small scale magnetic field properties and relating those to other ISM tracers. Comparing these results to observations that trace other ISM properties, including star formation and \HI\ line-width profiles, illuminates the roles played by both large- and small-scale magnetic fields in various dynamical processes within galaxies. More sophisticated and detailed models of the 3D magnetic structure \citep[such as those presented by][]{shneider_etal_2014} are a natural next step and indicate the sort of modeling that can be applied to large samples of galaxies observed with contiguous frequency coverage and excellent sensitivity. This new kind of investigation can be extended with the transformational capabilities of the SKA.

\subsection{Connection to properties of the ISM and star formation}

%

The ISM of galactic disks is highly turbulent and the recovery of its intrinsic polarised emission requires observations at high frequencies to avoid Faraday depolarisation (Beck et al., this volume). The Faraday depth spectra of turbulent fields contain many components that can be separated with very high resolution in real and in Faraday space \citep{frick_etal_2011}. The added value provided by the approach advocated in this chapter is the ability to also investigate field structures in the outer disks and halos of galaxies, where turbulence is reduced and polarised emission still significant. Observations of the same sample of galaxies at both high and moderate frequencies address complementary scientific goals and together probe the full galactic volume, on scales ranging from the smallest structure in the ISM to the galaxy as a whole. Are the small scale structures in the disk and halo the same, or are the turbulent properties decoupled?

\subsection{Large-scale field structures}

%

Depolarization has been crucial to understand the large-scale magnetic field structure in a sample of nearby galaxies \citep{heald_etal_2009,braun_etal_2010}. Due to the substantial depolarization in the midplane of star-forming galaxies \citep[e.g.][in the WSRT-SINGS survey and in the Virgo cluster, respectively]{braun_etal_2007,vollmer_etal_2013}, only the front layer of the ISM is visible in polarization as could be expected from internal Faraday dispersion \citep{sokoloff_etal_1998}. This depolarization, together with a common large-scale magnetic field topology, leads to a clear and consistent pattern within the sample: an azimuthal variation in fractional polarization exhibiting a minimum at the kinematically receding major axis, together with an azimuthal variation in the rotation measure. These characteristic patterns can be explained with a simple model for the large-scale magnetic field consisting of an axisymmetric trailing spiral together with a quadupolar configuration. Such configurations are predicted by e.g. the $\alpha-\Omega$ dynamo. The minimum in polarized intensity at the receding major axis vanishes when the same galaxies are observed at higher frequencies, where depolarization is reduced and the backside of the galaxy is visible. Additionally, faint polarized emission from the backside of the galaxy was also recovered at lower frequency in some galaxies \citep{braun_etal_2010}, having been substantially depolarized and with an additional dispersion term due to the intervening turbulent ISM. 

The SKA will have the capability to move beyond the recognition of this observational effect, and exploit the same physical properties of the magnetized ISM in galaxies to perform a more detailed form of Faraday tomography. When observing at higher frequency, the polarized synchrotron radiation originates from a somewhat greater depth in the ISM due to the frequency-dependent character of the depolarization. Making use of the variation in polarization and RM patterns, together with the field orientation, will lead to a more complete and accurate 3D picture of the large-scale magnetic field in galaxies and permit a more detailed model than the simplistic quadrupolar topology used by \citet{braun_etal_2010}. 

Very weakly polarized emission can be identified statistically in the output of RM Synthesis using the technique introduced by \citet{giessuebel_etal_2013}. The same strategy can be exploited with the SKA to study still weaker magnetic fields in the halos and outer disks of galaxies.




\subsection{Vertical field structures}

%

The structure of regular magnetic fields in the galactic plane has been obtained for a dozen or so different galaxies. The most important of these properties, in that they allow direct comparisons to be made with galactic dynamo theories, are the magnetic field pitch angles and rotational symmetry. The vertical distribution of both regular and random magnetic fields is much more poorly explored, even though these are also closely related to diverse dynamo theory predictions, as well as to other important aspects of galactic astrophysics such as gas flows across the disk-halo boundary. Measurements of how magnetic fields change as a function of height above the galactic plane has been difficult in the past due to the coarse resolution available in both spatial scales and in Faraday depth.

Important questions that will be resolved through study of the depolarization in galaxies with Band 2 of SKA1 include the following. What are the origins of halo magnetic fields: transport from a dynamo active disk, in-situ generation \citep[e.g.,][]{moss_etal_2010}, or a primordial leftover from galaxy formation? Are random magnetic fields transported into the halo, providing a route by which a disk dynamo can be saturated, and if so is this transport part of a self-limiting cycle involving the dynamo and outflows or just a fortuitous coincidence? Does a galactic wind/fountain transport magnetic fields out of the galaxy disk \citep[as observed by][based on an RM signature corresponding to a superbubble in NGC~6946]{heald_2012} and if so, are helical fields ejected into the halo \citep{heesen_etal_2011}? How does the magnetic energy density compare to the thermal energy density in outflows? Does magnetic reconnection contribute to gas heating in galaxy halos? How uniform are regular disk fields as a function of height above the galactic plane: in face-on galaxies are we measuring the mid-plane properties or an emission-weighted average and is the field structure the same through all $1$~kpc of the disk thickness or not? 

Faraday tomographic observations of a sample of disk galaxies, with a range of inclination angles, will allow us to establish, for the first time, three essential aspects of galactic magnetic fields that are key to firmly identifying their origin: (i) How the regular magnetic field structure changes with height above the disk, in particular do the components of the field have even or odd symmetry with respect to the mid-plane? (ii) What happens at the disk-halo boundary? Are small-scale and/or regular magnetic fields transported into the halo? (iii) Do the ordered fields in the disk and halo have similar or different patterns, and are X-shaped halos common? These field properties are all closely related to the state of the interstellar medium, in particular the buoyancy of hot gas, magnetic fields and cosmic rays driving vertical flows across the disk-halo boundary. The ability to resolve features at least on scales of 100~pc to 1~kpc is required to relate structures in the magnetic field to those in the gaseous distribution, such as \HI\ holes that have typical sizes smaller than 2~kpc \citep[e.g.,][]{bagetakos_etal_2011}. Moreover, typical correlation scales of magnetized structures in the ISM range from tens of pc to 1~kpc \citep[e.g.,][]{gaensler_etal_2005,fletcher_etal_2011,houde_etal_2013,mao_etal_2014}. Variations in the polarised intensity as a function of both Faraday depth and location in the plane of the sky will be compared and fit to the patterns expected from models with different symmetry properties in their magnetic field, in order to separate properties that are common in all galaxies from properties that are specific to individual galaxies.

\section{Synergy with other ISM observations}

%

A key aspect of the study of the Faraday thickness within galaxies is the connection between the magnetoionized structure implied by the depolarization spectra, and the gas phase properties determined from complementary observations. An important observable to relate to the Faraday structure of the ISM is the properties of the neutral gas traced by \HI.

Studies of the warm and cold \HI\ in other galaxies were already attempted by
\citet{younglo96, younglo97} and \citet{young2003} who analyzed \HI\
emission velocity profiles of seven nearby dwarf galaxies. By fitting
these profiles with Gaussian components, they found evidence for a
broad component with a velocity dispersion ranging from about 8 to
$13\,\mathrm{km\,s^{-1}}$ as well as a much narrower component with a velocity dispersion
ranging from 3 to $5\,\mathrm{km\,s^{-1}}$. The authors associated the narrow component
with the CNM and the broad component with the WNM phases of the ISM. A
similar study by \citet{dB6822} of the Local Group dwarf
galaxy NGC 6822 also found narrow and broad \HI\ components with
mean velocity dispersions of $4\,\mathrm{km\,s^{-1}}$ and
$8\,\mathrm{km\,s^{-1}}$,
respectively. They found that the narrow component is usually located
near star-forming regions, whereas the broad component tends to be
found along every line of sight. \citet{ianja2012} and \citet{stilp_etal_2013}
used stacking of velocity profiles to determine the fractions of cold and warm \HI. The key
limitation of these studies was angular resolution, and the improved
capabilities of the SKA should enable these studies to be repeated for individual profiles, rather than stacked ones. Faraday tomography will provide complementary information about the relevant coherence scales and degree of randomness in the ISM.

\section{Prospects with SKA}

%


\begin{table}
\begin{center}
\begin{tabular}{lcc}
Parameter & SKA1-MID & SKA1-SUR \\ \hline\hline
Lowest frequency (MHz) & 950 & 650 \\
Highest frequency (MHz) & 1760 & 1670 \\
RM resolution $\Delta\phi=2\sqrt{3}/\Delta\lambda^2$ (\radm) & 51.4 & 19.2 \\
Maximum RM scale $\pi/\lambda^2_\mathrm{min}$ (\radm) & 97.4 & 97.4 \\
Maximum RM$^a$ (\radm) & 5216 & 2443 \\ \hline
\end{tabular}
\caption{Parameters of interest in SKA1 Band 2 polarization observations.}
\label{table:summary}
\end{center}
\footnotesize{$^a$Assuming 1 MHz channel width; far higher RMs can be reached by forming individual polarization images with narrower bandwidth.}
\end{table}

As has been discussed elsewhere (e.g. Gaensler et al., this volume) the key technical aspect for polarization studies is the total $\lambda^2$ coverage. Here we assume that SKA1 Band 2 can be used in its entirety (Band 2 having been selected for the reasons outlined in \S\,\ref{section:science}).

In order to resolve small-scale structures in the disk and reduce beam (frequency-independent) depolarization, high angular resolution is required. For this frequency range and assuming SKA1-MID, $\nu_\mathrm{min}=950$ MHz. The current array configuration \citep{braun_2013,dewdney_2013} gives a 0.3 arcsec nominal resolution for a 200 km baseline length in SKA1-MID. However, such an extended configuration will not provide sufficient surface brightness sensitivity to reliably image the diffuse polarization in nearby galaxies. In order to match the surface brightness sensitivity obtained with existing instruments \citep[e.g. $10\,\mu\mathrm{Jy\,beam^{-1}}$ with $15^{\prime\prime}$ resolution;][]{braun_etal_2010}, with 100h of SKA1 observing time, an angular resolution of $1.5^{\prime\prime}$ is feasible. This is a dramatic step forward -- a factor of 10 reduction in synthesized beam size compared to existing instruments enables the study of magnetic field structures on $\approx34$~pc scales in the typical nearby grand-design galaxy M~83 \citep[$D=4.61$~Mpc;][]{saha_etal_2006}, sufficient to probe down to the correlation scale of the magnetic field structure -- but implies that only baselines out to $\approx40\,\mathrm{km}$ will usefully contribute to the observation. This is reflected in the sensitivity calculations in this chapter. In practice, for these calculations we excluded the outer 30 antennas in SKA1-MID, and the outer 18 antennas in SKA1-SUR. The MeerKAT and ASKAP arrays are assumed to remain in place and contribute to the sensitivities.

For proper analysis of the frequency-dependent depolarization in galaxies, the same angular resolution has to be considered at all frequencies. For this reason we have further adjusted the sensitivity calculation to reflect the fact that the outer antennas are effectively lost for the high-frequency end of the band. For both SKA1-MID and SKA1-SUR the effect is approximated by discarding the outer 10 antennas for the upper half of the frequency band, taking the SEFD values from the Science Performance document \citep{braun_2013}, and calculating image noise values with the formula provided in \S\,7.1 in that document.

\subsection{SKA Phase I}

Using the full available bandwidth, the broadband sensitivity from SKA1-MID band 2 after a deep 100h integration is estimated to be approximately $0.1\,\mu\mathrm{Jy\,beam^{-1}}$ (approximately 10 times more sensitive for an observation of fixed duration than the Jansky VLA at L-band), or about $1\,\mu\mathrm{Jy\,beam^{-1}}$ for a relatively shallow 1h observation. For SKA1-SUR, the same sensitivities are approximately 0.4 and $4\,\mu\mathrm{Jy\,beam^{-1}}$. The large sensitivity discrepancy is of obvious importance for single-field pointings that we consider for the study of individual galaxies; for a few galaxies with large angular size ($\gtrsim1^\circ$), multiple SKA1-MID pointings will be required, offsetting some of the benefit of the higher sensitivity provided by SKA1-MID. It should also be noted that in terms of RM precision, the larger bandwidth of SKA1-SUR does not necessarily provide a better observing capability, since the increased sensitivity of SKA1-MID compared to SKA1-SUR for a given observing time more than compensates for the larger $\lambda^2$ coverage of SKA1-SUR \citep[$\Delta\phi_\mathrm{eff}\approx\Delta\phi/\lbrack2\times S/N\rbrack$ by analogy to the astrometric precision in the image plane;][]{fomalont_1999}. In this case, the net improvement in effective RM precision is such that SKA1-MID has $\approx54\%$ better RM precision that SKA1-SUR, for the same on-source integration time. On the other hand, a narrower RMSF is always preferable to properly reconstruct closely spaced features in Faraday depth spectra \citep[features more closely spaced than the FWHM of the main lobe of the RMSF cannot be reliably interpreted; e.g.,][]{farnsworth_etal_2011,sun_etal_2014}. The use of SKA1-SUR may be preferable where recovery of complex Faraday depth spectra is judged to be more important than sheer RM precision.

SKA1-MID will be able to robustly study diffuse polarization at 100~pc resolution in galaxies out to 14~Mpc, and at 1~kpc resolution to $z\approx0.04$. The detailed study of magnetic field structures will be possible through shallow observations of hundreds of galaxies (so that similarities and peculiarities across the galaxy population can be investigated). Deeper polarization observations, ideally performed together with deep \HI\ observations, will enable an exquisite view of the connection between magnetism and ISM structure in tens of individual galaxies.

\subsection{SKA Phase I in its early stages}

During the ``early science'' period when each array has half of its final collecting area, excellent surface brightness sensitivity can still be achieved. Here we assume that the relative distribution of collecting area within the array configurations is held fixed, but the number of stations is halved. In that case the broadband sensitivity for a 100h observation in SKA1-MID is $160\,\mu\mathrm{Jy\,beam^{-1}}$, or $580\,\mu\mathrm{Jy\,beam^{-1}}$ for SKA1-SUR (both in Band 2 and using the entire bandwidth in each case). 

\subsection{SKA2}

For SKA2, the sensitivity to extended emission is far superior to current capabilities. Assuming ten times the sensitivity of SKA1-MID and using the same considerations as above, a 100h broadband integration would provide a sensitivity of $18\,\mathrm{nJy\,beam^{-1}}$ at the full angular resolution, assuming a 50\% sensitivity loss because of visibility weighting. SKA2 will enable Faraday tomography on 10~pc scales in nearby galaxies like M~83, with the same surface brightness sensitivity as existing facilities. Galaxies within 50~Mpc can be studied at 100~pc resolution. SKA2 will allow the extension of this study to substantially larger distances, permitting large samples and even opening the door to the tomographic study of the evolution of the 3D magnetic field structure in galaxies out to $z\approx0.125$, where structures on scales of $\approx1~\mathrm{kpc}$ can be resolved.

\section{Concluding Remarks}

Depolarization is an essential tool with which to study the three-dimensional structure of the magnetized ISM in galaxies. Both large- and small-scale magnetic structures will be revealed in exquisite detail by the SKA, even in its early incarnations. Detailed and sophisticated analysis techniques will be required to make the most of the transformational capabilities of the SKA. The SKA will provide a huge step forward in studying galactic magnetism: SKA1 represents an order of magnitude improvement in angular resolution for polarization studies of galaxies in the local Universe and an extension to a much larger volume than is feasible today, while SKA2 will extend the same study to a cosmologically interesting volume and illuminate the evolution of magnetic field structure in galaxies.


\bibliographystyle{apj}

\end{document}